\documentclass[prb,aps,amssymb,twocolumn,floatfix]{revtex4}

\usepackage{graphicx}
\usepackage{dcolumn}
\usepackage{bm}

\begin{document}

\preprint{}

\title{``Listening" to the spin noise of conduction electrons in bulk \emph{n}:GaAs}

\author{Scott A. Crooker$^1$, Lili Cheng$^1$, and Darryl L. Smith$^2$}

\affiliation{$^1$National High Magnetic Field Laboratory, Los Alamos
National Laboratory, Los Alamos, NM 87545}

\affiliation{$^2$Theoretical Division, Los Alamos National
Laboratory, Los Alamos, NM 87545}


\date{\today}
\begin{abstract}

We report a comprehensive study of stochastic electron spin
fluctuations -- spin noise -- in lightly doped ($n$-type) bulk GaAs,
which are measured using sensitive optical magnetometry based on
off-resonant Faraday rotation. Frequency spectra of electron spin
noise are studied as a function of electron density, magnetic field,
temperature, probe laser wavelength and intensity, and interaction
volume. Electron spin lifetimes $\tau_s$ are inferred from the width
of the spin noise spectra, and are compared with direct measurements
of $\tau_s$ using conventional Hanle effect methods. Both methods
reveal a strong and similar dependence of $\tau_s$ on the wavelength
and intensity of the probe laser, highlighting the undesired
influence of sub-bandgap absorption effects on the nominally
`non-perturbative' spin noise measurements. As a function of
temperature, the spin noise power increases approximately linearly
from 1.5~K to 30~K, as expected for degenerate electrons obeying
Fermi-Dirac statistics, but with an additional zero-temperature
offset. Finally, as the cross-sectional area of the probe laser
shrinks and fewer electrons are probed, the measured Faraday
rotation fluctuations due to electron spin noise are shown to
increase, as expected.
\end{abstract}

\maketitle
\section{Introduction and background}
Not long after the discovery of nuclear magnetic resonance, Felix
Bloch wrote in his seminal 1946 paper on \emph{Nuclear
Induction}\cite{Bloch} that ``Even in the absence of any orientation
by an external magnetic field one can expect in a sample with $N$
nuclei of magnetic moment $\mu$ to find a resultant moment of the
order $(N)^\frac{1}{2} \mu$ because of statistically incomplete
cancellation."  Thirty-nine years later these small, random
fluctuations within a nuclear spin ensemble -- spin noise -- were
directly observed by Sleator and co-workers\cite{Sleator} in a
low-temperature nuclear quadrupole resonance study of $^{35}$Cl
nuclei in NaClO$_3$. Subsequently, interest in spin noise phenomena
has been growing steadily, particularly in recent years as
experimental detection sensitivities continue to improve and as the
characteristic sizes of probed spin ensembles grow ever
smaller.\cite{MaminPRL, MaminPRB}  For example, proton nuclear spin
noise was measured in liquid samples at room
temperature,\cite{Gueron} and a theory of nuclear spin noise and its
detection was described \cite{Hoult}. More recently, spatial
distributions of nuclear spin noise have been imaged,\cite{Muller,
MaminNN} an important step towards an alternative and `passive'
approach to magnetic resonance imaging (MRI) that is based on a
system's intrinsic spin fluctuations alone.

In parallel with these efforts to detect nuclear spin noise,
experiments to measure the stochastic fluctuations of
\emph{electronic} spins have been pursued, first by Aleksandrov and
Zapassky \cite{Aleksandrov} who used optical Faraday rotation to
detect ground-state spin fluctuations in a gas of sodium atoms.
Within the last decade, related techniques to detect electronic spin
noise in atomic gases have been used to demonstrate spin squeezing
and also to control quantum-mechanical entanglement.\cite{Sorensen,
Kuzmich, Julsgaard, Geremia, Yabuzaki, Mitsui, Martinelli}

Recently, the frequency spectra of electron spin noise were
explicitly studied in classical (warm) vapors of rubidium and
potassium atoms.\cite{CrookerNature, Mihaila1} In accord with the
fluctuation-dissipation theorem, these noise signatures revealed the
full magnetic resonance spectrum of the atomic ground state, without
ever having to pump, excite, or otherwise perturb the spin ensemble
away from thermal equilibrium. These experiments also used an
off-resonant optical Faraday rotation probe to passively ``listen"
to the $\sqrt{N}$ spin fluctuations of the ensemble. The probe laser
in these studies was detuned by an energy $\Delta$ from an atomic
\emph{S-P} optical resonance, ensuring no absorption of the laser
(and therefore no perturbation of the atoms) to leading order.
Nonetheless, random spin fluctuations in the atomic ground state
imparted Faraday rotation fluctuations on the laser via the
dispersive (real) part of the vapor's dielectric function -- that
is, through the spin-dependent indices of refraction\cite{footnote1,
Happer} for circularly polarized light, $n^\pm$, which decay much
more slowly with laser detuning ($\sim$$\Delta^{-1}$) as compared to
the absorption ($\sim$$\Delta^{-2}$).

Similar optical approaches to measure electron spin noise in
condensed matter systems have now been demonstrated, notably in
electron-doped ($n$-type) GaAs by Oestreich and
co-workers.\cite{OestreichPRL, Romer}  These studies are especially
noteworthy in view of the rapidly-developing field of semiconductor
spintronics,\cite{Awschalom, Zutic} in that noise spectroscopy of
electron spins can reveal important dynamic spin properties (such as
spin relaxation time and precession phenomena) \emph{without}
needing to inject additional electrons by optical or electrical
means.  In this context, perhaps the simplest and most well-studied
system is the Fermi sea of spin-1/2 electrons that can form in the
conduction band of doped, direct-gap semiconductors such as GaAs.
While much about this spin system is known from extensive pump-probe
studies over the years, the spin noise properties of this ``ideal"
electron gas are only beginning to be explored.

To this end, this article reports on a comprehensive study of
stochastic electron spin noise in lightly electron-doped
(\emph{n}-type) bulk GaAs, which we measure using a sensitive
optical magnetometer based on sub-bandgap Faraday rotation.
Frequency spectra of electron spin noise are measured as a function
of electron density, applied transverse magnetic field, temperature,
probe laser wavelength and intensity, and interaction volume. We
infer electron spin lifetimes $\tau_s$ from the width of the spin
noise power spectra, and compare these values with direct
measurements of $\tau_s$ obtained using conventional methods based
on optical orientation of electron spins and the Hanle effect. Both
methods reveal a strong dependence of $\tau_s$ on the wavelength and
intensity of the probe laser, highlighting the undesired influence
of sub-bandgap absorption effects on these nominally
`non-perturbative' spin noise measurements. With decreasing
temperature from 30~K to 1.5~K, the noise power from this sea of
fluctuating electron spins decreases approximately linearly -- as
expected for degenerate electrons obeying Fermi-Dirac statistics --
but with an interesting zero-temperature offset. Finally, we show
that Faraday rotation fluctuations due to spin noise actually
increase as the area of the probe laser beam is reduced and fewer
electrons are probed.

\section{Experimental details}
Figure 1(a) shows a schematic of the optical magnetometer used to
passively `listen' to electron spin noise in $n$:GaAs. It is very
similar to that originally used to detect spin noise spectra in warm
vapors of alkali atoms.\cite{CrookerNature} A probe laser beam,
derived from a continuous-wave Ti:sapphire ring laser, is tuned to
the transparency region below the low-temperature band-gap of bulk
GaAs ($E_{gap}$$\sim$1.515 eV, or $\sim$818 nm). This probe laser is
linearly polarized and is focused through one of three silicon-doped
($n$-type) GaAs wafers that are mounted, strain-free, in the
variable-temperature insert of an optical $^4$He cryostat. Typical
probe laser spot diameters range from 15-150 $\mu$m.

The three bulk $n$:GaAs wafers (denoted A, B, and C) are
anti-reflection coated and are 350, 170, and 170 $\mu$m thick, with
electron densities $N_e$ = 1.4, 3.7, and 7.1 $\times 10^{16}$
cm$^{-3}$ at 10~K, respectively. These densities are near the
critical density at which the metal-insulator transition occurs in
$n$:GaAs ($N_e^{MIT} \simeq 2\times10^{16}$ cm$^{-3}$), where
electron spin lifetimes $\tau_s$ are known to be rather long at
cryogenic temperatures, of order 100 ns.\cite{Kikkawa, DzhioevPRB,
DzhioevPRL, FurisAPL, FurisNJP} A control wafer of semi-insulating
GaAs was also studied, and exhibited no detectable electron spin
noise signal.

Random fluctuations of the electron spins along the $\hat{z}$
direction, $\delta S_z (t)$, impart Faraday rotation fluctuations
$\delta \theta_F (t)$ on the transmitted probe laser beam via the
spin-dependent indices of refraction for right- and left-circularly
polarized light $n^\pm(\nu)$, as discussed in more detail in the
next Section. These Faraday rotation fluctuations are detected and
converted to a fluctuating voltage signal using a polarization
beam-splitter and a balanced photodiode bridge. We use either a 650
MHz bridge having 0.35 V/mW peak conversion gain (New Focus 1607),
or a slower 80 MHz bridge having 20 V/mW peak conversion gain (New
Focus 1807). The fluctuating voltage signals at the bridge output
are amplified and then detected using fast digitizers, similar to
the approach described recently by R\"{o}mer.\cite{Romer} Power
spectra of these time-domain signals are computed with
fast-Fourier-transform algorithms (using typical record lengths of
$2^{10}$ to $2^{15}$ points) and are signal-averaged in software.
Modest magnetic fields can be applied in the transverse direction
($B
\parallel \hat{x}$), which causes all spin fluctuations $\delta S_z$
to precess about $B_x$. This shifts the peak of the spin noise away
from zero frequency (where other environmental noise sources may
exist), to the electron Larmor precession frequency $\omega_L=g_e
\mu_B B_x/\hbar$, from which the electron \emph{g}-factor, $g_e$,
can be measured ($\mu_B$ is the Bohr magneton).

\section{A short theoretical description}
Faraday rotation -- the optical polarization rotation of linearly
polarized light upon passage through a material -- results from
unequal indices of refraction for right- and left-circularly
polarized light, $n^\pm(\nu)$:
\begin{equation}
\theta_F(\nu) = \frac{\pi \nu L}{c} [n^+(\nu) - n^-(\nu)]
\end{equation}
where $\nu$ is the frequency of the light, $c$ is the speed of
light, and $L$ is the effective thickness of the material. In
analogy with noise spectroscopy of alkali
atoms,\cite{Aleksandrov,CrookerNature} a difference between
refraction indices $n^+(\nu)$ and $n^-(\nu)$ arises near the
band-edge of GaAs when the the net spin polarization of electrons in
the conduction band is not zero. (The coupling between electron/hole
spin orientation and circular optical polarization is given by the
well-known optical selection rules in GaAs and related
semiconductors,\cite{OO} which in turn derive from spin-orbit
splitting in the valence band.)

In the absence of a magnetic field along the laser direction
$\hat{z}$,  $n^+ - n^-$ scales with the difference between spin-up
and spin-down electron densities, $N_e^+ - N_e^-$ (where the total
electron density is $N_e = N_e^+ + N_e^-$, and `spin-up' and `-down'
denote electrons with spin projection antiparallel and parallel to
$\hat{z}$). For photon energies $h\nu$ well below the GaAs band-edge
at $h\nu_0$ (the latter being where absorption changes due to spin
imbalances mainly occur), the energy dependence of the index
\emph{difference} $n^+(\nu) - n^-(\nu)$ can be approximated using
Kramers-Kronig relations to scale inversely with laser detuning,
$\Delta = \nu_0 - \nu$. Thus for large detuning,
\begin{equation}
n^+(\nu) - n^-(\nu) \sim \frac{1}{\Delta}(N_e^+ - N_e^-).
\end{equation}
Using Eq.~(2) we can now capture generally how the magnitude of the
detected spin noise depends on various external parameters. We will
explicitly measure the dependence of $n^+(\nu) - n^-(\nu)$ on
detuning $\Delta$, using a fixed electron spin polarization, in
Section IV-D.

In these noise studies, the number of electrons $N$ within a probe
laser beam of cross-sectional area $A$ and over the sample thickness
$L$ is $N$=$N_e AL$. At zero magnetic field and in thermal
equilibrium, this ensemble of $N$ electrons has zero time-averaged
spin polarization: $\langle N^+ - N^- \rangle$=0. Electron spin
noise, however, arises from statistical temporal fluctuations in the
quantity $N^+ - N^-$, which have root-mean-square (rms) amplitude
\begin{equation}
\sqrt{\langle(N^+ - N^-)^2\rangle}=\sqrt{fN} = \sqrt{fN_eAL}.
\end{equation}
Here, the factor $f$ accounts for the \emph{fraction} of the
electron spins that are allowed to fluctuate. For non-interacting
electron spins such as those found in the warm (classical) alkali
vapors studied previously\cite{CrookerNature} or, \emph{e.g.}, in
the case of dilute paramagnetic impurities in a
solid,\cite{MaminPRL} all $N$ electrons fluctuate and $f$=1. In
contrast, for degenerate electron systems obeying Fermi-Dirac
statistics, only those electron spins within thermal energy
$\sim$$k_B T$ of the Fermi energy $\epsilon_F$ have available phase
space to fluctuate (all states at lower energy being occupied), in
which case $f<1$. For an ideal Fermi sea of electrons and in the
absence of other correlations,\cite{Mihaila2} $f \rightarrow 0$ as
$T \rightarrow 0$. These considerations will be discussed in Section
IV-E, where the temperature dependence of the electron spin noise is
measured. Our $n$:GaAs samples, being lightly doped near the
metal-insulator transition, are neither clearly in the low-doping
limit (where electrons are localized on isolated donors and can be
considered non-interacting), nor clearly in the high-doping limit
(where $\epsilon_F$ greatly exceeds the donor binding energy and
Fermi-Dirac statistics of degenerate electrons dominate).

Combining Eqs. (1)-(3) and ignoring overall constants, the rms
amplitude of Faraday rotation fluctuations due to electron spin
noise in $n$:GaAs therefore scales as
\begin{equation}
\sqrt{\langle\theta_F^2 \rangle} \sim \frac{1}{\Delta}
\sqrt{\frac{L}{A}} \sqrt{fN_e}.
\end{equation}
This total spin noise -- in units of radians of measured Faraday
rotation -- should therefore scale approximately inversely with
laser detuning (measured explicitly in Section IV-D), and inversely
with the square root of the probe beam diameter (measured in Section
IV-F).

\begin{figure}[tbp]
\includegraphics[width=.48\textwidth]{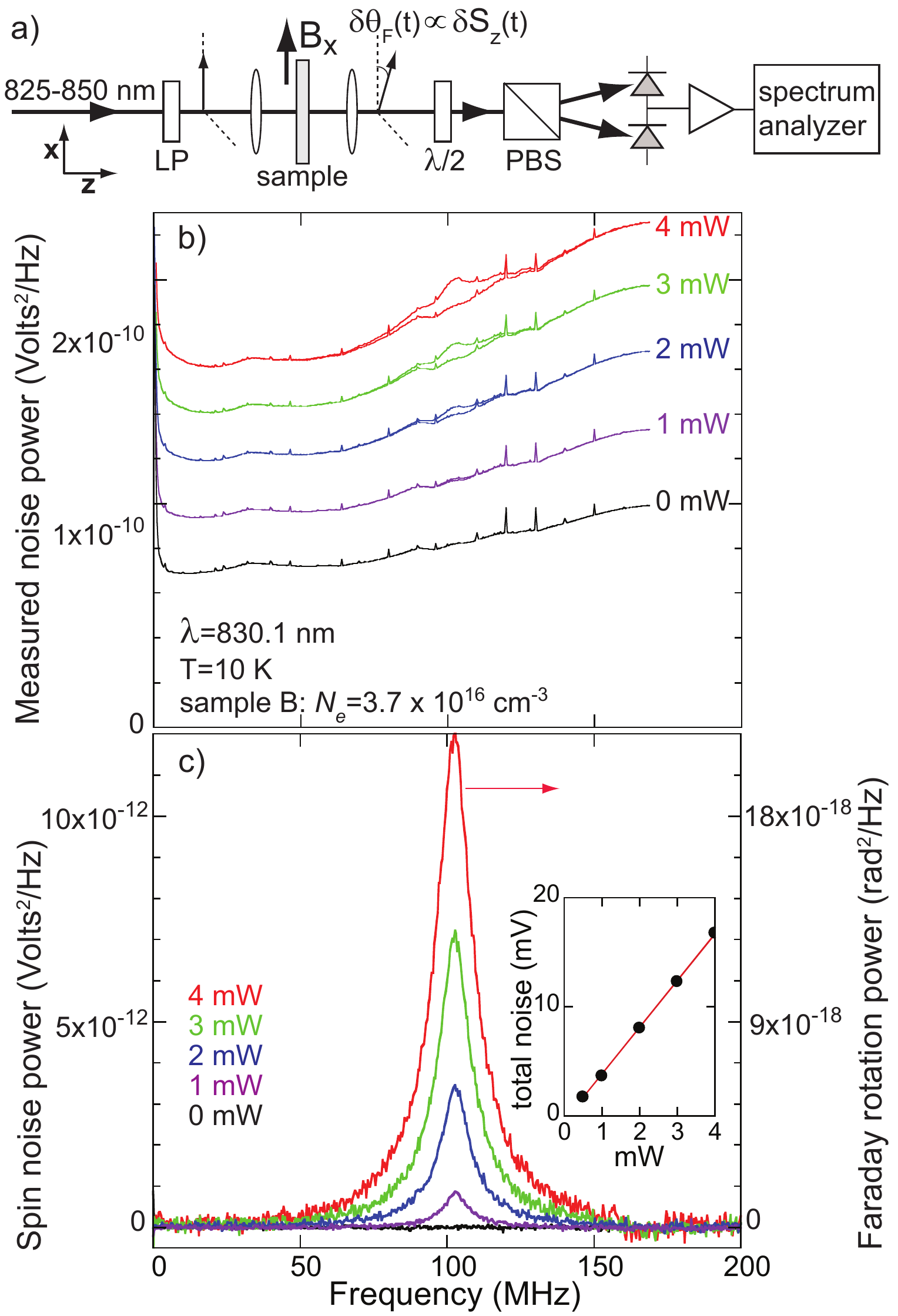}
\caption{a) The spin noise experiment, showing linear polarizer
(LP), polarization beamsplitter (PBS), and half-wave plate
($\lambda$/2). b) Raw spin noise data from an $n$:GaAs wafer (sample
B), for transmitted probe laser intensities from 0-4 mW. Data are
not offset; the increasing noise power density arises from
increasing photon shot noise. At each laser intensity, two spectra
are acquired: one in the target transverse magnetic field (here,
$B_x$=175 G), and one in a background field ($B_x$$>$1000 G). c)
Their difference reveals the extra noise power density due to
fluctuating, precessing electron spins, shown in units of measured
volts$^2$/Hz. The 4~mW spectrum is also expressed as a Faraday
rotation power density (radians$^2$/Hz; right axis). Inset: The
integrated spin noise (in units of volts, or square-root of the
integrated power) scales linearly with laser intensity. All these
spectra have the same total integrated Faraday rotation noise:
$\sim$22.5 $\mu$radians.} \label{fig1}
\end{figure}

\section{Spin noise measurements}
\subsection{Dependence on probe laser intensity}
An example of raw data from an electron spin noise experiment on
\emph{n}:GaAs is shown in Fig.~1(b). Here, the temperature of sample
B is 10~K, the probe laser wavelength is tuned below-gap to 830.1
nm, and the intensity of the probe laser beam at the output of the
sample -- that is, the transmitted laser intensity -- is varied from
0-4 mW. (Note that in this paper we refer to the probe laser power
as an ``intensity", so as to avoid potential confusion with the
``noise power" that we measure). For each probe laser intensity,
noise power spectra at two transverse magnetic fields are acquired:
one at the target field ($B_x$=175 G in this case), and one at a
large background field (typically, $B_x$$>$1000 G) that shifts the
spin noise out of the detected frequency range. The difference
between these two power spectra [shown in Fig.~1(c)] reveals any
extra noise power due to the probed ensemble of randomly fluctuating
and precessing electron spins in the $n$:GaAs.

Unless otherwise stated, we measure and show spectra of the measured
noise \emph{power} density -- that is, in units of (volts)$^2$/Hz of
detected signal, or more usefully (since voltages vary trivially
with detector and amplifier gains) in units of (radians)$^2$/Hz of
detected Faraday rotation.  Frequency-integrated (or total) spin
noise -- see Eq.~(4) -- is computed from the measured noise power
spectra, and is expressed either as a total spin noise power
($\langle \theta_F^2 \rangle$, in units of radians$^2$), or simply
as a total spin noise ($\sqrt{\langle \theta_F^2 \rangle}$, in units
of radians).

In Figure 1(b), the noise power spectra at 0 mW -- when the probe
laser is turned off -- reveals the background electronic noise floor
of the amplified 650 MHz photodiode bridge output.  Sharp features
at specific frequencies are due to insufficiently-shielded nearby
radiofrequency sources. The increase of the background noise floor
with increasing probe laser intensity is due to additional ``white"
photon shot noise.  Using $\sim$3 mW of transmitted probe laser, the
photon shot noise power density is comparable to the electronic
noise power density of these detectors. At about the same probe
intensity, the extra noise due to fluctuating electron spins
(at $\sim$100 MHz) becomes visible on this scale. Clearly,
the spin noise signals from electrons in $n$:GaAs are smaller
in terms of absolute noise power density than the background noise
floor, and signal averaging of several minutes to an hour is typically
required.

The spin noise power spectra are much more clearly seen in the
difference spectra of Fig.~1(c), for which the detector and photon
shot noise contributions are subtracted away. The spin noise power
spectra exhibit Lorentzian lineshapes, indicating that the spin-spin
correlation function, $\langle S_z(t) S_z(0) \rangle$, decays
exponentially with characteristic spin relaxation time $\tau_s$. The
full-width at half-maximum of the spectral peaks, $\Gamma$,
therefore reveals the inverse electron spin lifetime:
\begin{equation}
\tau_s = 1/(\pi \Gamma)
\end{equation}

The integrated area under these spectral peaks yields the total
measured spin noise power, in units of volts$^2$ or radians$^2$. For
a fixed peak width $\Gamma$ and a fixed averaging time, the
visibility of the spin noise peaks (or ratio of spin
noise:background noise) improves with increasing probe laser
intensity. As shown in the inset of Fig.~1(c), doubling the probe
intensity doubles the measured voltage fluctuations that are due to
spin noise (and quadruples the measured power), while the background
voltage density due to photon shot noise increases only by
$\sqrt{2}$.  Put another way, the integrated Faraday rotation
induced by spin fluctuations is independent of probe laser intensity
(\emph{e.g.}, $\sqrt{\langle \theta_F^2 \rangle} \simeq
22.5~\mu$radians for all the noise spectra in Fig.~1(c)), but the
Faraday rotation noise floor due to photon shot noise decreases as
the inverse square root of the probe intensity.

\begin{figure}[tbp]
\includegraphics[width=.37\textwidth]{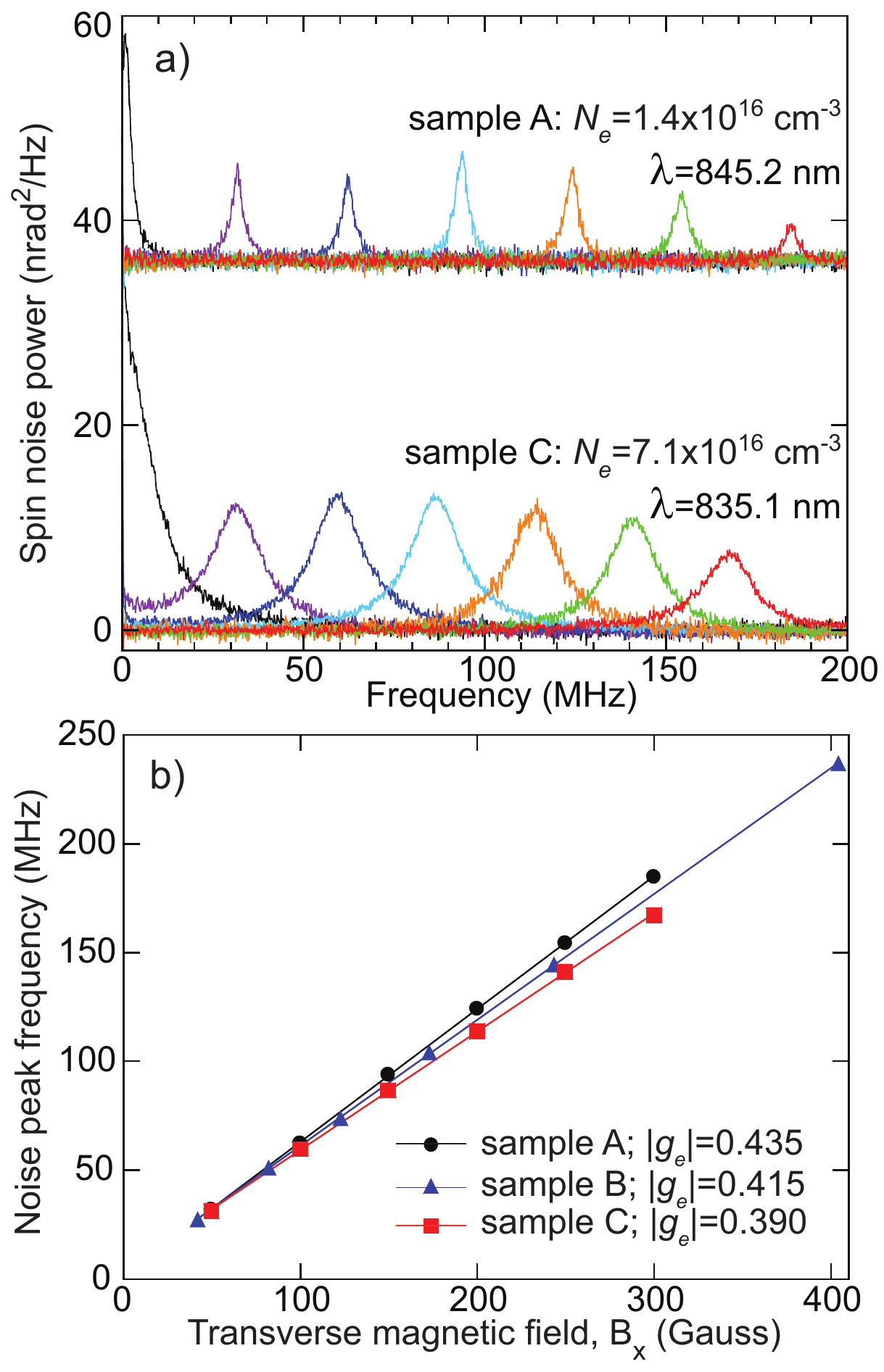}
\caption{a) Electron spin noise power spectra at 10~K from
\emph{n}:GaAs wafers A and C, at $B_x$=0, 50, 100, 150, 200, 250,
300 G (black$\rightarrow$red). b) The center frequency of these spin
noise peaks versus $B_x$ for all three \emph{n}:GaAs wafers. The
slope reveals the electron $g$-factor, $|g_e|$.} \label{fig2}
\end{figure}

\subsection{Dependence on transverse magnetic field, $B_x$}

Figure 2(a) shows a series of electron spin noise power spectra from
$n$:GaAs wafers A and C, in the presence of applied transverse
magnetic fields $B_x$ from 0 - 300~G.  The background noise floor
from the detectors and from photon shot noise has been subtracted.
The spin noise peaks shift to higher frequencies with increasing
$B_x$ as expected from the electron Larmor precession frequency,
$\omega_L = g_e \mu_B B_x/\hbar$. However the two series of noise
peaks do not shift at precisely the same rate. This can be more
clearly seen in Fig.~2(b), which plots the spin noise frequency as a
function of $B_x$ for all three $n$:GaAs samples. The different
slopes reveal the different low-temperature electron $g$-factors in
the three samples, which are found to decrease slightly in magnitude
as the electron density (and therefore the Fermi energy
$\epsilon_F$) increases, in reasonable agreement with the
established\cite{Hopkins} energy dependence in GaAs,
$g_e(\epsilon_F)$= -0.44 + 6.3 eV$^{-1}$ $ \times \epsilon_F$.

Figure 2(a) also reveals that the widths $\Gamma$ of the spin noise
peaks from sample A are considerably narrower than those from sample
C (4.5~MHz versus 16~MHz), implying a longer electron spin lifetime
$\tau_s$ ($\sim$70 ns versus $\sim$20 ns). While this relationship
ultimately proves to be true for samples A and C, it should not
strictly be inferred from the data shown in Fig.~2(a) -- the next
Section discusses how $\tau_s$ can be adversely influenced
(\emph{i.e.}, reduced) by external factors such as probe laser
wavelength, intensity, and spot size. Indeed, under very weak probe
conditions sample A exhibits noise peaks narrower than 1.8~MHz at
10~K, indicating that $\tau_s >$ 175~ns.

It is also apparent from these raw data that the area under the
noise peaks varies slightly with $B_x$.  However, these variations
should not be considered significant here, as no attempt was made to
correct for the frequency-dependent gain of the photodiodes, the
amplifiers, or the digitizers (especially at high frequencies, where
bandwidth-limiting filters attenuate incoming signals). Nor is it
significant that in Fig.~2(a) the integrated spin noise power from
sample C is larger than that from sample A. The experimental
parameters were very different when these two samples were measured:
not only were the probe laser wavelengths different (845.2 nm for
sample A versus 835.1 nm for sample C), but the sizes of the focused
laser spots were different and the sample thicknesses themselves
were different. As shown in Eq.~(4), all these experimental
parameters directly influence the total detected spin noise.

\subsection{Measuring electron spin lifetimes $\tau_s$ using spin noise spectroscopy}

The intrinsic spin lifetimes $\tau_s$ of conduction band electrons
in $n$:GaAs can, in principle, be inferred from the widths $\Gamma$
of spin noise power spectra [see Eq.~(5)].  In the limit that the
probe laser itself does not perturb the electronic states in the
semiconductor (meaning, essentially, that no probe laser photons are
absorbed), then spin noise spectroscopy represents a passive and
non-perturbative probe of time-dependent electron spin correlations.
That is, spin dynamics are revealed through their stochastic
fluctuations alone and no optical pumping or intentional optical
orientation of electron spins is required, in contrast to most
conventional pump-probe studies of electron spin dynamics which
necessarily perturb the electron spin ensemble away from thermal
equilibrium.\cite{Kikkawa, DzhioevPRB, DzhioevPRL, FurisAPL,
FurisNJP} As suggested previously,\cite{CrookerNature, OestreichPRL}
non-perturbative approaches based on spin noise may prove
advantageous for studying the dynamics of electron spin systems at
low temperatures, or (especially) for probing systems containing few
spins, as demonstrated recently in the context of magnetic resonance
force microscopy.\cite{MaminPRL, MaminPRB, Rugar}

In practice, however, we find that optical spin noise spectroscopy
as applied to electrons in bulk $n$:GaAs can be significantly
influenced by the undesired effects of probe laser absorption, even
when probing well below the low-temperature bandgap of GaAs. In
contrast to optical spin noise spectroscopy of alkali
vapors\cite{CrookerNature} (which have sharp, atomic absorption
resonances), it is considerably more challenging in bulk
semiconductors to operate the probe laser in a regime that is
clearly ``non-perturbative", while still retaining sufficient signal
to measure.  In large part this is due to the long, low-energy,
exponentially-decaying absorption tail (``Urbach tail") that is
characteristic of bulk semiconductors. As shown below, measurements
of $\tau_s$ via spin noise spectroscopy can be adversely influenced
by the residual sub-bandgap absorption of the probe laser itself.
Accurate measurements of $\tau_s$ are shown to require either very
little probe laser intensity (in which case the spin noise signals
are small, as shown in Fig.~1), or very large wavelength detunings
below the GaAs band-edge (in which case the noise signals are also
very small, as per Eq.~(4) and as also studied in the next Section).

\begin{figure}[tbp]
\includegraphics[width=.47\textwidth]{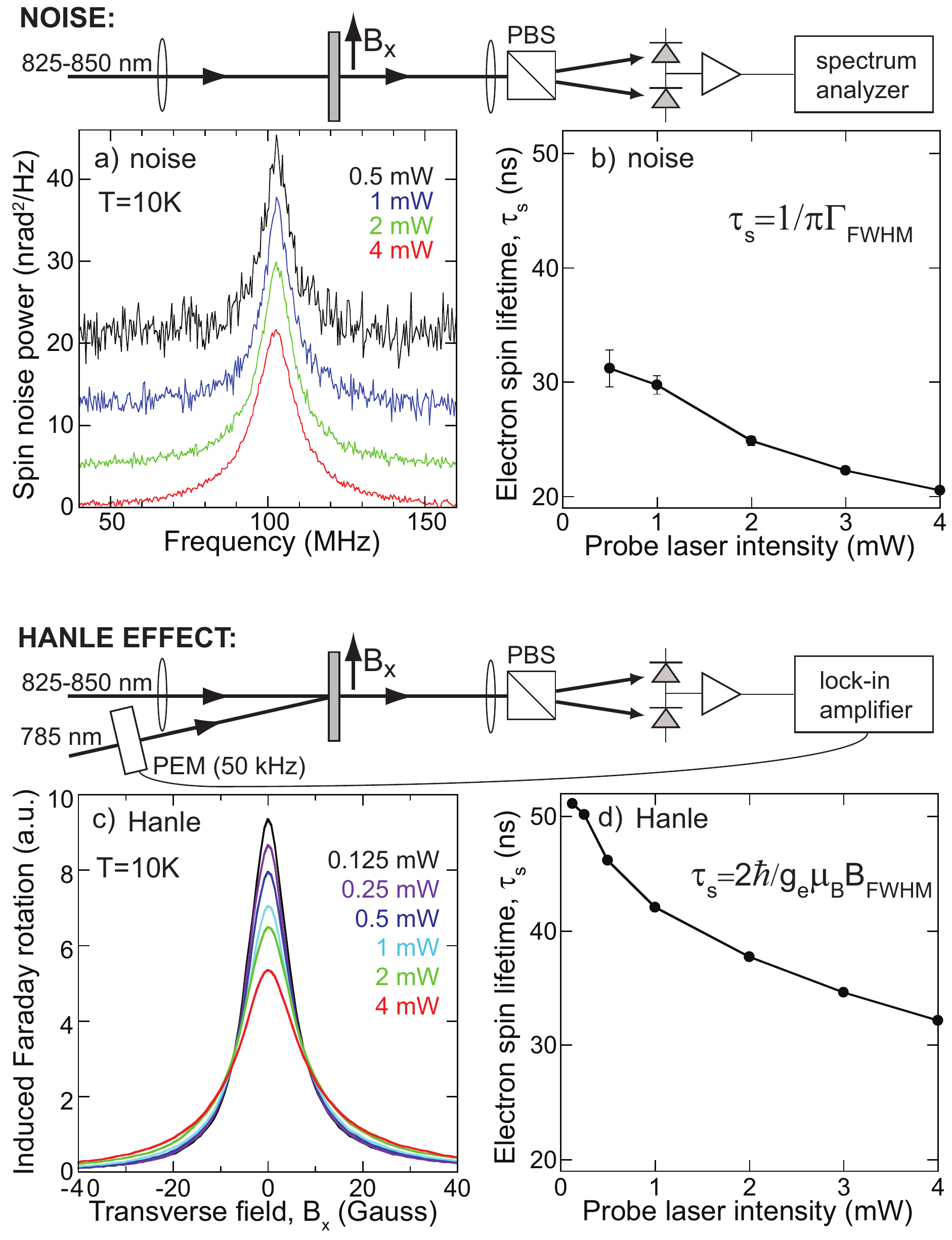}
\caption{a) The electron spin noise power in sample B at 10~K using
transmitted probe laser intensities of 0.5, 1, 2, and 4 mW
($\lambda$=830.1 nm; spectra offset for clarity). b) The spin
lifetime $\tau_s$ inferred from the width $\Gamma$ of the noise
spectra decreases with increasing probe laser intensity. c)
Measuring $\tau_s$ in sample B via optical spin orientation and
`conventional' Hanle-effect methods, using probe intensities from
0.125 to 4.0 mW ($\lambda$=830.1 nm) d) The inferred $\tau_s$ from
these Hanle data also decreases with increasing probe laser
intensity.} \label{fig3}
\end{figure}

To illustrate these points, Figs. 3(a,b) show $\tau_s$ measured in
sample B at 10~K using spin noise spectroscopy. For reference, a
schematic of the spin noise experiment is also shown. With the probe
laser wavelength at 830.1~nm, four noise power spectra are acquired
using transmitted probe laser intensities of 0.5, 1, 2, and 4 mW.
The spin lifetime $\tau_s$ inferred from the width $\Gamma$ of these
spectra are shown in Fig.~3(b).  Clearly $\tau_s$ is not constant,
but rather decreases with increasing probe laser intensity,
indicating that absorption effects are adversely influencing the
measurement.

To provide a direct comparison, we also measure $\tau_s$ using
``conventional" methods based on \emph{intentional} optical
orientation of electron spins and the Hanle effect.  This
experimental setup, sketched above Fig.~3(c), is identical to the
setup for spin noise spectroscopy except that electron spins in the
$n$:GaAs wafer are now partially aligned along $\pm \hat{z}$ by an
additional, above-bandgap (1.58 eV), defocused pump laser. The beam
path and 25 $\mu$m spot size of the probe laser on the sample is
identical for the two methods.  The polarization of the pump laser
is modulated by a photoelastic modulator from left- to
right-circular at 50~kHz, optically orienting electron spins
parallel or antiparallel to $\hat{z}$. This small and constant
electron spin polarization, $S_z$, imparts Faraday rotation
$\theta_F$ on the probe laser at this frequency, which is detected
with lock-in amplifiers. Applied transverse magnetic fields $B_x$
depolarize the injected spins by an amount that depends on $\tau_s$,
leading to a reduction of the induced signal [see Fig.~3(c)] -- this
is the basis of the Hanle effect,\cite{OO} which is routinely used
to measure $\tau_s$ in GaAs and other semiconductors. These Hanle
curves exhibit characteristic Lorentzian line shapes,
$\theta_F(B_x)\propto 1/[1+(g_e \mu_B B_x \tau_s / \hbar)^2]$, with
full-widths $B_{fwhm}=2\hbar/g_e \mu_B \tau_s$ from which the
effective spin lifetime is revealed. We verify that we operate in
the weak-pumping regime, where $\theta_F$ scales linearly with (and
$\tau_s$ is independent of) pump laser intensity. Figure 3(c) shows
a series of Hanle curves from sample B at 10~K, where the probe
laser intensity is increased from 0.125~mW to 4~mW. The spin
lifetimes extracted from these Hanle curves are shown in Fig.~3(d),
where it is clear that -- even though $S_z$ is constant -- $\tau_s$
decreases with increasing probe laser intensity, similar to the
trend exhibited by the spin noise measurements in Fig.~3(b).

\begin{figure}[tbp]
\includegraphics[width=.4\textwidth]{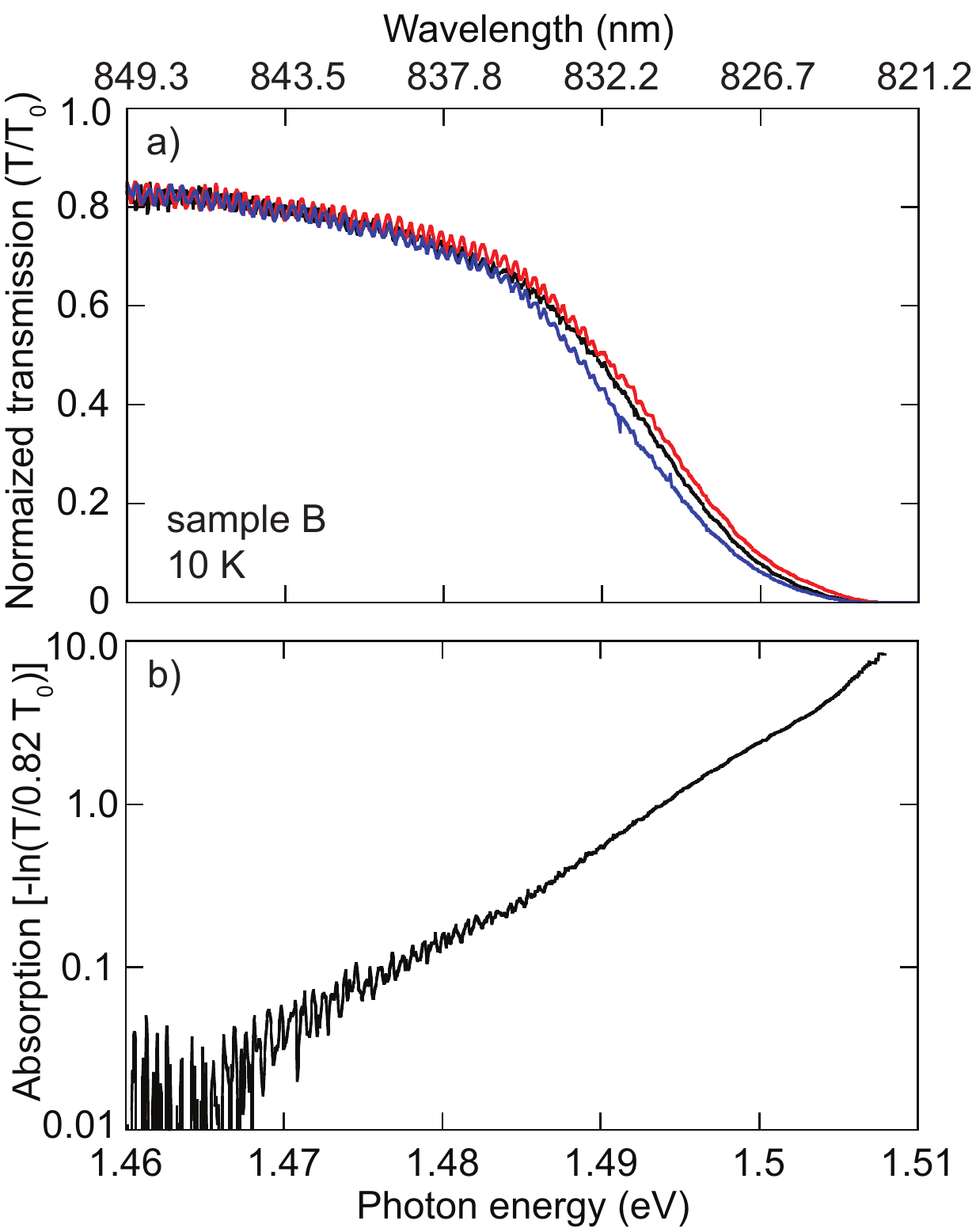}
\caption{a) The normalized optical transmission spectrum, T/T$_0$,
through \emph{n}:GaAs sample B at 10~K, using different probe laser
intensities (the 10~K bandgap of GaAs is $\sim$1.515 eV). b) The
same data plotted as an optical absorption (and assuming 18\%
reflection by the sample). The Urbach absorption tail persists well
below the nominal GaAs bandgap. Absorption at these sub-gap energies
can reduce the electron spin lifetime that is measured by spin noise
spectroscopy.} \label{fig4}
\end{figure}

To understand these results it is essential to independently measure
the absorption and transmission characteristics of these $n$:GaAs
wafers. Figure 4(a) shows the normalized transmission of the probe
laser through sample B at 10~K, for photon energies from 1.46~eV up
to near the GaAs bandgap at $\sim$1.515~eV. The blue, black, and red
traces were acquired using transmitted probe laser intensities of
0.04, 0.60, and 1.80 mW at 832 nm (1.49 eV). Differences between
these traces arise from absorption and self-bleaching of the probe
laser as it passes through the sample. The probe transmission is
zero near the band-edge (where absorption is strong), and increases
to about 82\% when the laser is tuned well below the band-edge. The
transmission does not saturate near 100\% at low photon energies,
likely due to some reflection of the probe laser by the Si$_3$N$_4$
coating (similar behavior was observed from all of our coated
$n$:GaAs wafers).  Regardless, Fig.~4(a) indicates that sizeable
absorption exists even at energies well below the GaAs band-edge. To
see this exponentially-decaying absorption tail more clearly, Fig.
4(b) shows this data expressed as an optical absorption constant
($\alpha L$) and plotted on a semi-log scale (and assuming 18\%
reflection).

Using these data it is possible to show that the changes in
$\tau_s$, as measured either by spin noise spectroscopy or by
conventional Hanle-effect methods, are directly correlated with the
amount of probe laser intensity that is \emph{absorbed} by the
$n$:GaAs wafer, independent of the actual probe laser wavelength or
intensity.  Figure 5 shows a compilation of data from samples B and
C where $\tau_s$ was measured by both spin noise spectroscopy and
also by the Hanle effect, using a variety of probe laser wavelengths
and intensities.  When $\tau_s$ is plotted as a function of the
absorbed probe laser intensity, all the points collapse onto a
common curve. The strong reduction of $\tau_s$ in the regime of
large absorption very likely results from faster electron spin
relaxation due to the density of photoexcited holes, by to the
Bir-Aronov-Pikus (BAP) spin relaxation mechanism.\cite{BAP} The
apparent spin lifetime increases by up to a factor of three in these
studies as we approach the ``non-perturbative" regime wherein few
probe laser photons are absorbed, either by tuning to very long
wavelengths ($\lambda >$ 845 nm) or by using very low probe laser
intensity, or both. Note that increasing the laser spot size has a
similar effect (not shown), as this also reduces the photoexcited
hole density. In the limit of small probe laser absorption, the
measured $\tau_s$ approaches its intrinsic upper bound at 10~K
($\sim$60 ns for sample B, and $\sim$30 ns for sample C), which is
limited in these $n$-type samples by Dyakonov-Perel spin
relaxation.\cite{OO}

\begin{figure}[tbp]
\includegraphics[width=.40\textwidth]{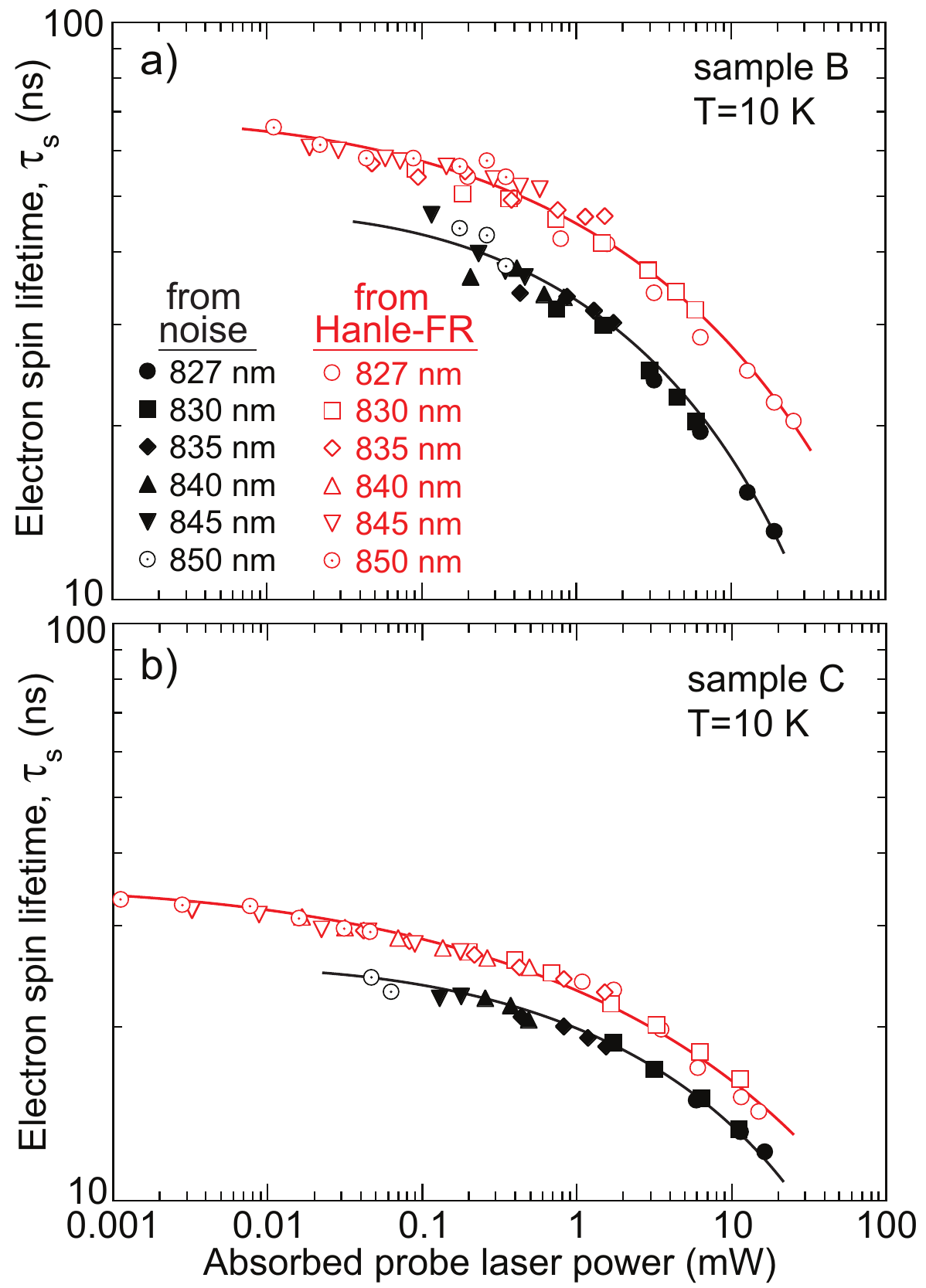}
\caption{a) A direct comparison of the electron spin lifetime
$\tau_s$ measured by spin noise methods (black points) and by
`conventional' Hanle-effect methods (red points), in \emph{n}:GaAs
sample B at 10~K. $\tau_s$ is shown on a log-log scale as a function
of the probe laser intensity that is \emph{absorbed} by the sample.
When plotted in this way, data from experiments using various probe
laser wavelengths and intensities collapse onto common curves (lines
are guides to the eye). For both methods, $\tau_s$ decreases
significantly with absorbed probe laser intensity, likely due to
electron-hole creation by the probe laser itself. b) A similar
comparison from \emph{n}:GaAs sample C.} \label{fig5}
\end{figure}

While these two methods exhibit very similar trends as a function of
absorbed probe intensity, they also reveal that $\tau_s$ as measured
by spin noise spectroscopy is consistently 30-40\% shorter than
$\tau_s$ measured by conventional Hanle methods. In principle, both
methods should yield similar $\tau_s$.  However, we note that
noise-based spin lifetime studies of mobile spins are susceptible to
transit-time effects, wherein the spins being measured diffuse out
of the probed volume in a characteristic time that is shorter than
the true spin relaxation time. Transit-time broadening effects are
well known in studies of atomic gases\cite{Kominis} (where spin
lifetimes and therefore spin diffusion lengths are very long), and
may be important here in lightly-doped bulk $n$:GaAs because the
characteristic spin diffusion length of electrons in these
samples\cite{FurisNJP} is of order 10 $\mu$m -- comparable to the 25
$\mu$m diameter of the focused laser spot.  Noise measurements using
larger spot diameters ($>$50 $\mu$m) were found to yield longer
$\tau_s$ values, in better agreement with $\tau_s$ measured by
Hanle-effect methods.

Comparing the $\tau_s$ measurements from spin noise spectroscopy and
from the Hanle effect, it is clear that both methods are equally
susceptible to the undesired effects of probe laser absorption.
Moreover, we find that -- at least for bulk $n$:GaAs -- a
``non-perturbative" regime (in which the intrinsic $\tau_s$ is
accurately measured) is more easily achieved using conventional
Hanle-effect methods. That is, both the pump and probe lasers can
readily be made sufficiently weak so as not to adversely influence
$\tau_s$, and the measurements continue to exhibit very good
signal-to-noise within a few minutes' time [note the good
signal-to-noise of the data in Fig.~3(c), even when using very low
probe laser intensity]. In contrast, we find that spin noise
spectroscopy using similar probe wavelengths and intensities
requires considerably more signal-averaging.  We note that
Hanle-effect methods based on photoluminescence\cite{DzhioevPRB,
DzhioevPRL} or magneto-optical Kerr effects\cite{FurisAPL, FurisNJP}
have been used in recent years to measure some of the longest spin
lifetimes in $n$:GaAs -- in excess of 500 ns in some cases.

\subsection{Dependence on probe laser detuning, $\Delta$}

The magnitude of Faraday rotation fluctuations due to spin noise,
$\sqrt{\langle \theta_F^2 \rangle}$, is expected to follow the
energy dependence of the refraction index difference, $n^+(\nu) -
n^-(\nu)$, as outlined earlier in Section III. In simple atomic
systems having nearly ideal Lorentzian absorption resonances,
$n^+(\nu) - n^-(\nu)$ decays inversely with probe laser detuning
$\Delta$ when the spin polarization is finite.\cite{Happer,
CrookerNature} Faraday rotation fluctuations due to spin noise in
alkali vapors were verified\cite{CrookerNature} to decrease as
$\Delta^{-1}$, confirming that spin fluctuations coupled to the
`passive' optical probe primarily through the dispersive indices of
refraction and not through any absorptive effect.

To make a similar comparison in $n$:GaAs, where the band-edge
absorption is considerably less idealized, the decay of $n^+(\nu) -
n^-(\nu)$ at energies below the band-edge must be measured directly
(the $\Delta^{-1}$ scaling used in Eq.~(2) was approximate, for the
purposes of outlining general trends). Thus, here we aim to measure
$\sqrt{\langle \theta_F^2 \rangle}(\nu)$, the energy dependence of
Faraday rotation fluctuations due to spin noise in $n$:GaAs, and
directly compare it to the independently-measured decay of $n^+(\nu)
- n^-(\nu)$ in the presence of a small and fixed spin polarization.

Figure 6(a) shows a series of spin noise power spectra obtained from
sample B at 10~K, where the probe laser wavelength was tuned below
the $n$:GaAs band-edge from 830.6~nm to 850.4~nm
(small$\rightarrow$large detuning). The transmitted probe laser
intensity was maintained at 3~mW, and the probe spot size and sample
temperature were fixed. Clearly, electron spin noise induces larger
Faraday rotation fluctuations as $\Delta$ is reduced. The blue
square points in Fig.~6(b) show the integrated spin noise
$\sqrt{\langle \theta_F^2 \rangle}$ under these spectra (in
microradians), as a function of probe laser photon energy.

\begin{figure}[tbp]
\includegraphics[width=.4\textwidth]{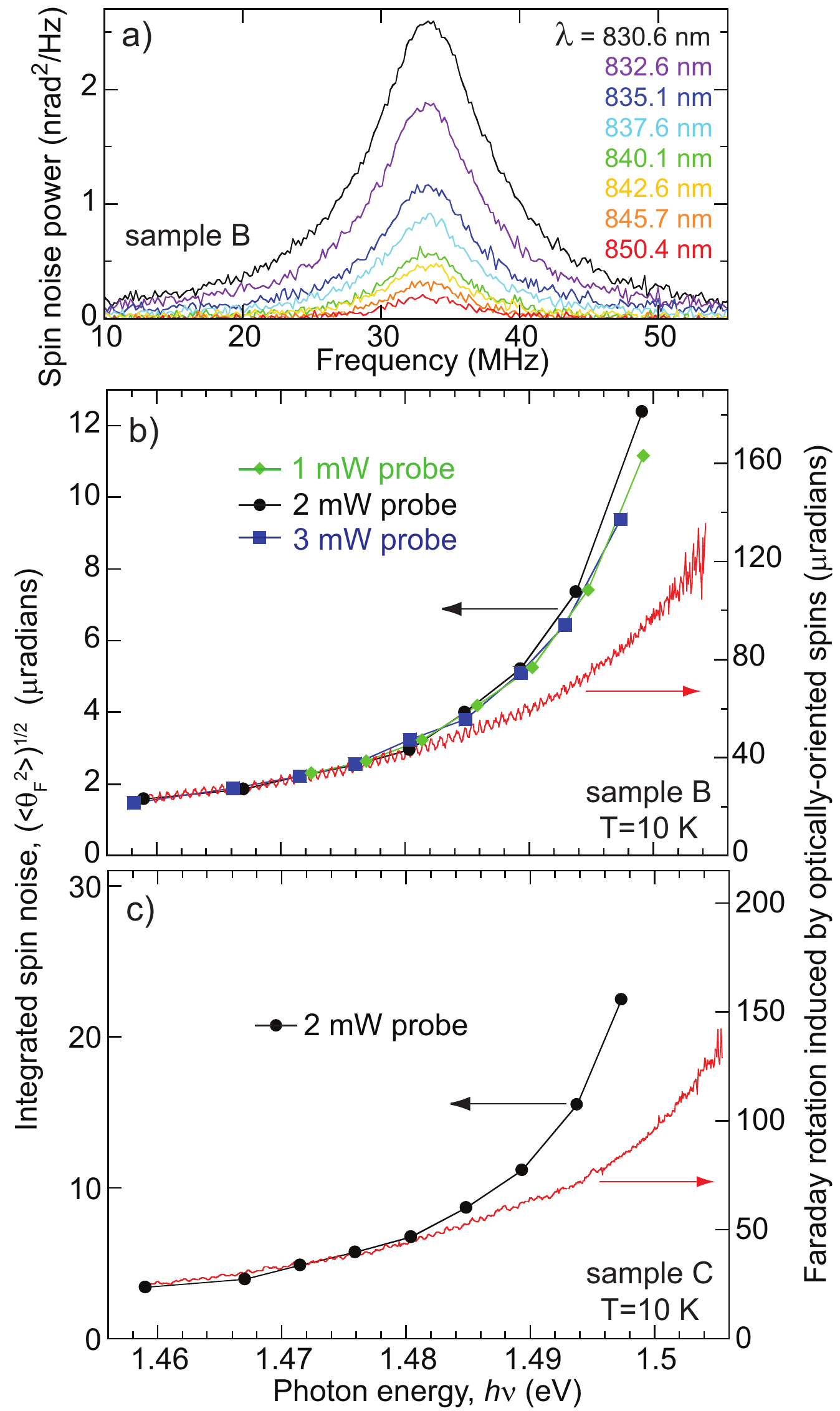}
\caption{a) Electron spin noise power spectra from \emph{n}:GaAs
sample B at 10~K, using sub-bandgap probe laser wavelengths from
830.6~nm to 850.4~nm. For all spectra, the transmitted probe laser
intensity was 3 mW, and $B_x$=57 G. b) The integrated electron spin
noise ($\sqrt{\langle \theta_F^2 \rangle}$, in $\mu$radians) versus
probe laser photon energy $h\nu$, for three probe laser intensities.
For comparison, the continuous red trace (right axis) shows
$\theta_F (\nu)$ induced by intentional optical orientation of
electron spins in sample B, using an additional 1.58 eV
circularly-polarized pump laser. $\theta_F (\nu)$ decays as
$\sim$$\Delta^{-1}$, assuming a GaAs band-edge at 1.515 eV. c) A
similar comparison from sample C.} \label{fig6}
\end{figure}

To interpret these data and to provide an accurate comparison, we
independently measure $n^+(\nu) - n^-(\nu)$ in sample B. This is
accomplished by measuring the Faraday rotation $\theta_F$ that is
induced on the probe beam by a small and constant electron spin
polarization $S_z$ that is \emph{intentionally} injected into the
sample using an additional, above-bandgap, defocused and
circularly-polarized pump laser.  This measurement uses the same
experimental Hanle-effect setup described in the previous Section
(and depicted in Fig.~3), but with $B_x$=0 and with continuous
tuning of the probe laser wavelength. This small injected spin
polarization perturbs the spin densities $N_e^\pm$ in the $n$:GaAs
wafer, and therefore modifies the associated indices of refraction
$n^\pm(\nu)$ by a constant amount. The Faraday rotation of the
transmitted probe laser therefore measures explicitly the photon
energy dependence of $n^+(\nu)-n^-(\nu)$, which is shown by the
continuous red curve in Fig.~6(b) (right axis). These studies were
performed in the weak-pump and weak-probe limit, where $\theta_F$
scaled linearly with pump intensity and was independent of probe
intensity. Assuming a GaAs band-gap at 1.515 eV and fitting the red
curve to a power law, $n^+(\nu)-n^-(\nu)$ does indeed decay very
nearly as $\Delta^{-1}$ (the fitted exponent is -1.06).

Figure 6(b) therefore directly compares the energy dependencies of
$\sqrt{\langle \theta_F^2 \rangle}$ (solid points) and $n^+(\nu) -
n^-(\nu)$ (red curve). The agreement between the two is reasonable
at large detunings, \emph{i.e.}, at photon energies below
$\sim$1.48~eV ($\lambda > 838$ nm). For smaller detunings, the
dependencies diverge -- Faraday rotation fluctuations due to spin
noise increase more rapidly than the Faraday rotation induced by a
small, fixed spin polarization.

While it is tempting to attribute this disparity to absorption
effects and associated electron gas heating (the measured spin noise
\emph{does} increase with temperature, as discussed in the next
Section), repeated studies using different probe laser intensities
do \emph{not} exhibit any systematic changes. As shown in Fig.~6(b),
nearly identical results were obtained using 1, 2, or 3 mW of
transmitted probe laser (green, black, and blue points), suggesting
that absorption effects are not adversely influencing the total spin
noise. It may be that the different spin polarization profiles of
the two methods plays a role: whereas the fluctuating spin
polarization that gives rise to spin noise exists throughout the
entire wafer, the intentionally injected spin polarization $S_z$ is
generated only within a spin diffusion length of the $n$:GaAs
surface. However, possible surface effects have not been explicitly
investigated in this work. Similar results were obtained in all the
$n$:GaAs samples -- Fig.~6(c) shows the results of a similar
comparison in sample C.

\subsection{Dependence on temperature}

The integrated spin noise that we measure in $n$:GaAs scales as the
square root of the number of fluctuating electron spins. As
discussed earlier [see Eqs.~(3) and (4)], this number may represent
only a fraction $f$ of the \emph{total} number of electrons $N$ if,
being fermions, the electrons form a degenerate system and
Fermi-Dirac statistic apply. In this case, only the electron spins
within the thermal energy $k_B T$ of the Fermi energy $\epsilon_F$
have available phase space to fluctuate; all states at lower
energies are fully occupied and spin fluctuations are suppressed. In
an ideal electron gas, the number of fluctuating electrons $fN$ can
be estimated as
\begin{equation}
fN=V\int^\infty_0 f(\epsilon)[1-f(\epsilon)]g(\epsilon)d\epsilon,
\end{equation}
where $V$ is the probed sample volume, $f(\epsilon)$ is the
Fermi-Dirac distribution using the appropriate
(temperature-dependent) chemical potential, and $g(\epsilon)$ is the
density of states in the conduction band of bulk $n$:GaAs ($\propto
\sqrt{\epsilon}$). Therefore, $fN$ is expected to be constant at
high temperatures where the gas is classical and the electrons are
non-interacting ($k_B T \gg \epsilon_F$; $f \sim 1$), and is
expected to decrease when a degenerate electron gas forms upon
cooling. At very low temperatures ($k_B T < \epsilon_F$), $fN$ is
expected to decrease linearly to zero as $T \rightarrow 0$.

\begin{figure}[tbp]
\includegraphics[width=.40\textwidth]{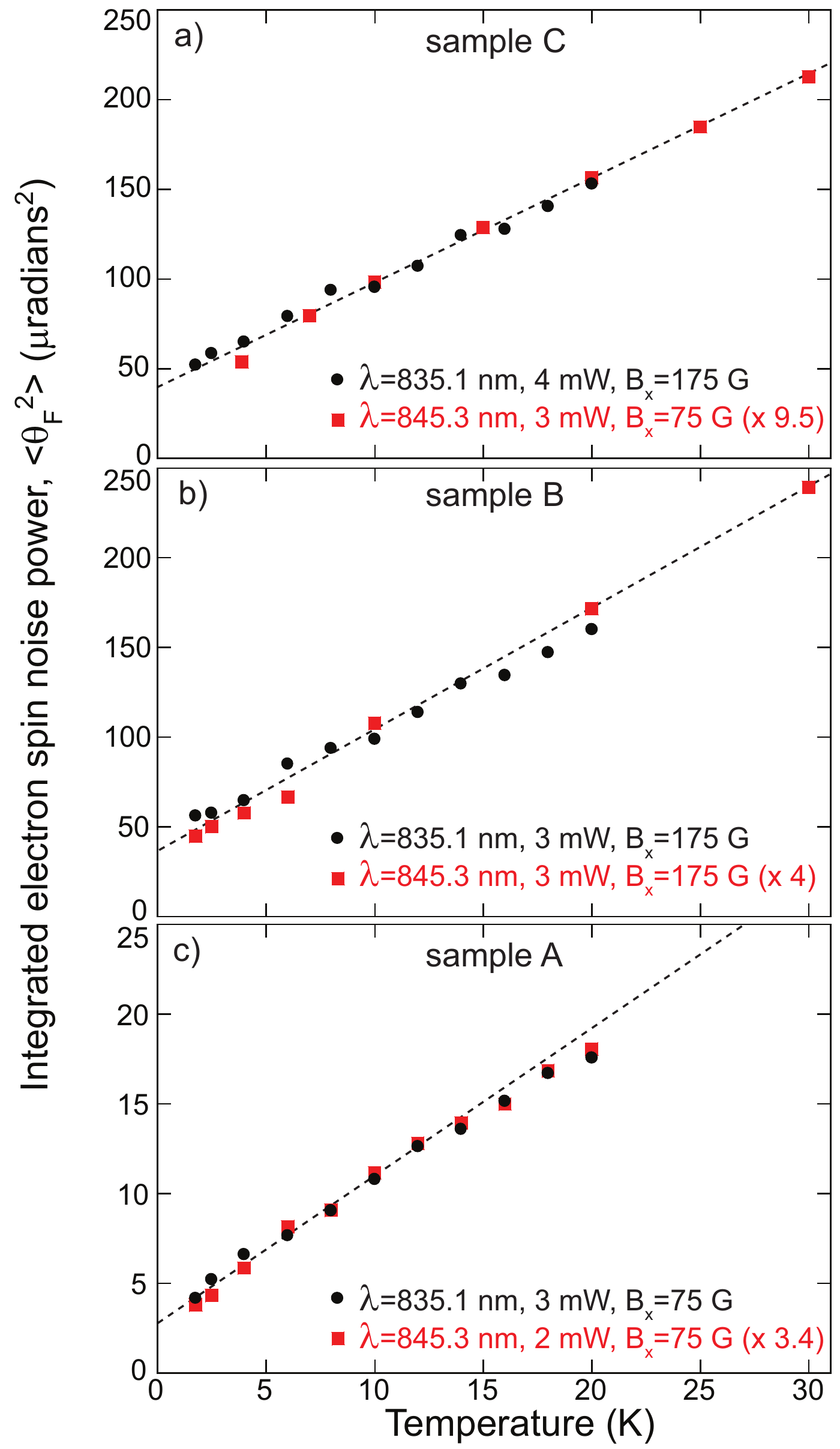}
\caption{a) The temperature dependence of the integrated electron
spin noise \emph{power} ($\langle \theta_F^2 \rangle$, in units of
$\mu$radians$^2$) from \emph{n}:GaAs sample C. Temperature sweeps
corresponding to red and black points used different probe laser
wavelengths, intensities, and spot sizes, but are scaled so as to
overlap. b,c) Similar temperature dependencies from \emph{n}:GaAs
samples B and A.  Dotted lines are linear guides to the eye.}
\label{fig7}
\end{figure}

The measured temperature dependence of the spin noise \emph{power}
$\langle \theta_F^2 \rangle$, which scales with $fN$, is shown in
Fig.~7 for all three $n$:GaAs wafers, from $T$=30~K down to
$T$=1.5~K. For all three samples, the measured noise power does
indeed decrease approximately linearly as $T \rightarrow 0$,
indicating that electrons in these $n$:GaAs samples do form
degenerate electron systems to which Fermi-Dirac statistics apply.
However, the most striking aspect of these data -- observed in all
three samples -- is that $\langle \theta_F^2 \rangle$ does
\emph{not} appear to intercept the origin when the data are
extrapolated to zero temperature. Rather, a finite amount of
electron spin noise remains as $T \rightarrow 0$.

To ensure that inadvertent heating or absorption effects played no
significant role, the temperature dependence of $\langle \theta_F^2
\rangle$ was measured multiple times on each sample using different
probe laser intensities, wavelengths, and spot sizes. To within
overall scaling constants, the same temperature dependencies were
observed, regardless of experimental conditions. The characteristic
``noise" on these data may be inferred from the scatter of the data
points, and derives primarily from intensity drifts of the probe
laser.  In comparison with prior studies,\cite{Romer} we do not
observe any discontinuities or non-monotonic behavior in the
temperature dependence of $\langle \theta_F^2 \rangle$.

This zero-temperature offset in the measured spin noise power is
puzzling, but may derive from the fact that these $n$:GaAs wafers
have electron densities $N_e$ that are in the range of the critical
metal-insulator transition density $N_e^{MIT} \simeq 2 \times
10^{16}$ cm$^{-3}$. That is, these $n$:GaAs wafers are neither
clearly in the high-doping regime (where the system is a good metal
and Fermi-Dirac statistics overwhelmingly dominate), nor are they
clearly in the low-doping regime (where electrons are localized and
non-interacting). For very low electron densities $N_e \ll
N_e^{MIT}$,  where the system is best viewed as an ensemble of
isolated and non-interacting electrons localized on their respective
donors, Fermi-Dirac statistics are not expected to apply and all
electrons are expected to fluctuate, giving a constant spin noise
that is independent of temperature. Very approximately, then, the
trends observed in Fig.~7 may therefore result from the combined
influence of electrons that are best described as mostly-localized,
and electrons that are best described as mostly-free. Measurements
of spin noise in $n$:GaAs having significantly larger or smaller
$N_e$ should help elucidate these findings.

\subsection{Dependence on probe laser spot size}

An often-mentioned advantage of noise-based spin measurements is the
favorable scaling of spin noise signals with decreasing system
size.\cite{MaminPRL,Muller} In an ensemble of $N$ spins, the ratio
of measured spin noise to the measured signal from a fully
magnetized ensemble ($\sim$$\sqrt{N}/N$) necessarily increases as
$N$ is reduced. In practice, it has been demonstrated that
$\sqrt{N}$ spin fluctuations can already exceed the thermal
equilibrium (``Boltzmann") magnetization of $N$ paramagnetic spins
in typical applied fields, when $N$ (or equivalently, when the
sample volume) is small.\cite{MaminPRL, MaminPRB} Indeed,
noise-based techniques were the basis for the recent detection of
\emph{single} electronic spins using ultrasensitive cantilevers
\cite{Rugar} or scanning tunneling microscopes\cite{Durkan, Messina,
Nussinov} and noise techniques will likely continue to provide the
basis for detection of nuclear spin resonance in nanometer-scale
structures. \cite{Muller, MaminNN}

\begin{figure}[tbp]
\includegraphics[width=.40\textwidth]{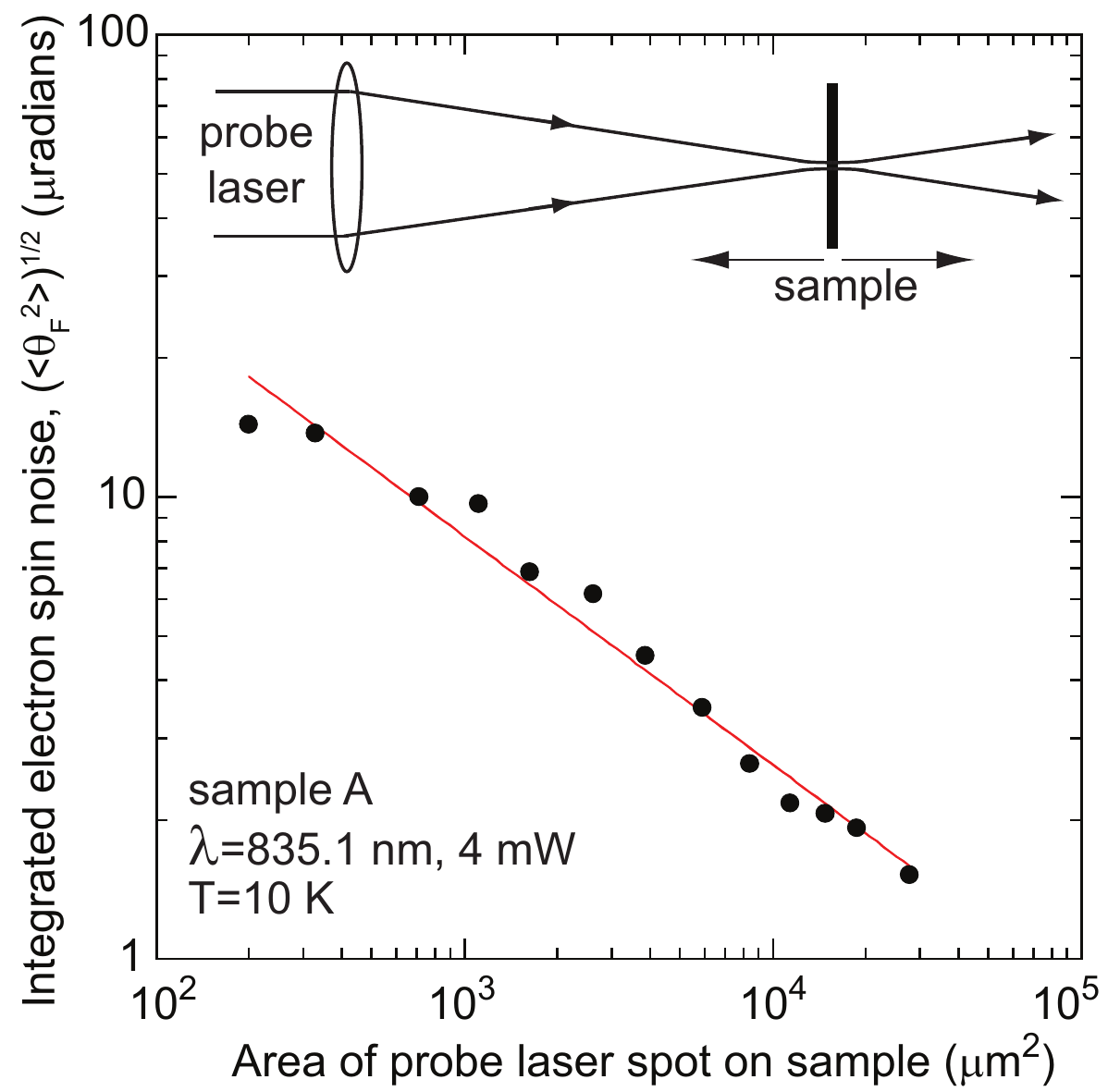}
\caption{Log-log plot of the integrated Faraday rotation
fluctuations due to electron spin noise ($\sqrt{\langle \theta_F^2
\rangle}$, in units of $\mu$radians) in \emph{n}:GaAs sample A as a
function of the cross-sectional area of the probe laser spot on the
sample. The line shows a 1/$\sqrt{\textrm{area}}$ dependence, as
expected for spin noise [see Eq.~(4)]. Inset: The laser spot size
was varied by translating the sample near the probe laser focus, and
assuming Gaussian optics.} \label{fig8}
\end{figure}

For fluctuating electron spins in $n$:GaAs, the advantageous scaling
of noise signals with shrinking system size can be directly
investigated by reducing the cross-sectional area of the probe laser
beam in the sample, as shown in Fig.~8. As the probe beam shrinks,
the total integrated noise $\sqrt{\langle \theta_F^2 \rangle}$
increases as $1/\sqrt{\textrm{area}}$, in good agreement with
Eq.~(4). (Concurrently, the number of probed electrons $N$ decreases
from $\sim$$10^{11}$ to $\sim$$10^9$).

That not only the relative magnitude but the \emph{absolute}
magnitude of $\sqrt{\langle \theta_F^2 \rangle}$ increases with
shrinking interaction volume is a consequence of the Faraday
rotation detection method. An absolute increase in spin noise signal
is not expected, for example, in conventional magnetometers that
employ pickup coils. Consider a fully-polarized spin ensemble: The
Faraday rotation $\theta_F$ imparted to a transmitted laser beam
depends only on the areal density of spins $N_e L$ and not on the
total number of probed spins $N$ [see Eq.~(2)], and $\theta_F$ is
therefore independent of the beam's cross-sectional area. Therefore,
the effective sensitivity of the measurement -- defined as the
Faraday rotation per polarized spin, $\theta_F /N$ -- is larger in
smaller beams that probe fewer spins.  Spin fluctuations, which
scale as $\sqrt{N}$, therefore induce correspondingly more Faraday
rotation when using smaller beams.

\section{Summary}

In bulk $n$:GaAs, spin noise spectroscopy using sub-bandgap Faraday
rotation revealed the dynamical properties of conduction electron
spins ($\tau_s, g_e$) through their fluctuation spectra alone, in
keeping with the fluctuation-dissipation theorem. Spin noise spectra
were studied as a function of electron concentration, magnetic
field, temperature, probe laser wavelength and intensity, and sample
volume. On the balance, these measurements indicated that the
integrated area, frequency, and width of these spin noise spectra
were in reasonable agreement with a simple model [Eq.~(4)].

However, systematic studies also made clear that this optical
approach to spin noise spectroscopy -- at least as applied to bulk
$n$:GaAs -- is by no means a panacea.  These noise methods are
susceptible to undesired absorption of the probe laser, even in the
nominal transparency region below the GaAs band-edge (Figs.~3-5).
Absorption effects ultimately lead to incorrect values of the
intrinsic electron spin lifetime $\tau_s$, unless care is taken to
operate the probe laser in a regime that is demonstrably
``non-perturbative" (long wavelengths $\lambda > 845$ nm, low
intensities, and/or large spot sizes), in which case the spin noise
signals are very small.  We found, moreover, that accurate
measurements of $\tau_s$ are more readily and quickly obtained using
conventional techniques based on the Hanle effect and intentional
optical orientation. Nonetheless we posit that these limitations may
be somewhat relaxed in cleaner or lower-dimensional semiconductor
structures that have less-pronounced absorption tails.

Some puzzles remain: The zero-temperature offset observed in the
temperature dependence of the spin noise (Fig.~7) is against simple
expectations of an ideal Fermi gas, but may arise in these bulk
$n$:GaAs samples from the localizing influence of the embedded
silicon donors. Also, the divergence between the energy dependencies
of the spin noise and the refraction index difference at small laser
detuning (Fig.~6) is not understood at this time.

Nonetheless, the outlook for optical spectroscopy of electron spin
noise in semiconductors is promising. Clear signatures of conduction
electron spin noise are measurable in $n$:GaAs under a variety of
conditions, and they reveal important dynamical information about
the spins themselves. The favorable scaling of spin noise signals
with reduced system size (Fig.~8) suggests its use for studying spin
correlations in mesoscopic electron ensembles, as very recently
reported.\cite{GMMuller} And finally, recent proposals for spin
noise spectroscopy using time-delayed pairs of ultrafast laser
pulses can potentially extend measurable noise bandwidths out to
terahertz frequencies,\cite{Starosielec} while application of
oscillating magnetic fields may permit the observation of
multi-photon phenomena in the spin noise.\cite{Braun}

\section{Acknowledgements}

We thank Shanalyn Kemme at Sandia National Lab for
antireflection-coating the GaAs wafers, and we are indebted to
Bogdan Mihaila, Peter Littlewood, and Sasha Balatsky for valuable
discussions. We acknowledge support from the Los Alamos LDRD program
and the National High Magnetic Field Laboratory.

\end{document}